\documentclass[reqno]{amsart}
\def\be{\begin{equation}}
\def\ee{\end{equation}}
\def\ba{\begin{array}}
\def\ea{\end{array}}
\def\bea{\begin{eqnarray}}
\def\eea{\end{eqnarray}}

\begin{document}


\title[Ettore Majorana and his heritage seventy years later]{Ettore
Majorana and his heritage \\ seventy years later}

\author[S. Esposito]{Salvatore Esposito}
\address{{\it S. Esposito}: Dipartimento di Scienze Fisiche,
Universit\`a di Napoli ``Federico II'' \& I.N.F.N. Sezione di Napoli, Complesso
Universitario di M. S. Angelo, Via Cinthia,
80126 Napoli ({\rm Salvatore.Esposito@na.infn.it})}%

\begin{abstract}
The physicists working in several areas of research know quite
well the name of Ettore Majorana, since it is currently associated
to fundamental concepts like {\it Majorana neutrinos} in particle
physics and cosmology or {\it Majorana fermions} in condensed
matter physics. But, probably, very few is known about other
substantial contributions of that ingenious scholar, and even less
about his personal background. For non specialists, instead, the
name of Ettore Majorana is usually intimately related to the fact
that he disappeared rather mysteriously on March 26, 1938, just
seventy years ago, and was never seen again.
\\
The life and the work of this Italian scientist is the object of
the present review, which will also offer a summary of the main
results achieved in recent times by the historical and scientific
researches on his work.
\end{abstract} 

\maketitle

\section{Introduction}

\begin{quote}
Some time after joining the Fermi group, Majorana already had such an erudition and
reached such a high level of comprehension of physics that he was able to discuss with
Fermi about scientific problems.  Fermi himself held him to be the greatest theoretical
physicist of our time.  He often was astounded [...].  I remember exactly these words
that Fermi spoke: `Once a problem has already been posed, no one in the world is able of
solving it better than Majorana'.
\end{quote}

Ettore Majorana's fame\footnote{This review of Majorana's life and
work is mainly based on work by the author published elsewhere
(see below for references). For biographical information about
Majorana, see Ref. \cite{Recami}.} rests on testimonies like this
one by Bruno Pontecorvo \cite{Pontecorvo}, a younger colleague of
Majorana at the Institute of Physics in Rome directed by Enrico
Fermi. It can be promptly recognized just by borrowing Fermi's own
words, given in 1937 in the occasion of a meeting of the board in
charge of the competition for a new full professorship in
theoretical physics in Italy.\footnote{Note, however, that Fermi
expressed similar opinions about Majorana also in several other
occasions; see Ref. \cite{Recami}.}

\begin{quote}
  Without listing his works, all of which are highly notable
both for their originality of the methods as well as for the importance of the results
achieved, we limit ourselves to the following:

  In modern nuclear theories, the contribution made by this
researcher to the introduction of the forces called ``Majorana forces" is universally
recognized as the one, among the most fundamental, that allows us to understand
theoretically nuclear stability.  The work of Majorana today serves as a basis for the
most important research in this field.

  In atomic physics, the merit of having resolved some of the
most intricate questions on the structure of spectra through
simple and elegant considerations of symmetry is due to Majorana.

  Lastly, he devised a brilliant method that deals with
positive and negative electron in a symmetric way, eventually eliminating the necessity
to rely on the extremely artificial and unsatisfactory hypothesis of an infinitely large
electrical charge diffused in space, a question that had been tackled by many other
scholars without success.\footnote{This and many other interesting documents may be found
(unfortunately only in Italian) in the comprehensive book in Ref. \cite{Recami}.}
\end{quote}

With this justification, the board, chaired by Fermi, proposed to apply for the second
time a special bill passed few years early in order to give a chair to the Nobel Prize
Guglielmo Marconi, and suggested to the Minister of National Education ``to appoint
Majorana as full professor of Theoretical Physics at some University of the Italian
kingdom, for high and well-deserved repute, independently of the competition rules''. The
Minister accepted the proposal: evidently, such a ``reputation'' was sufficiently
established on the basis of just few (nine) papers published by the Italian scientist.

Unfortunately enough, the University of Naples hosted his talent for three months only,
until the end of March 1938, when Majorana gave his last lesson.

As recalled by one of his students, Gilda Senatore,\footnote{See what reported in Ref.
\cite{Moreno} and references therein.} on Friday March 25, the day after his 21st lesson,
``differently than what he usually did [when no lecture on theoretical physics was
scheduled], Majorana came to the Institute and stayed there for few minutes. From the
corridor leading to the small room where I was writing, he called me by name: `Miss
Senatore...'; he didn't enter in the room but remained in the corridor; I reached him and
he gave me a closed folder telling: `here's some papers, some notes, keep them... we will
talk about later'; afterward he went away and, turned back, said again 'we will talk
about later'. The day after, the director of the Institute of Physics at the University
of Naples, Antonio Carrelli, received a mysterious telegram from Palermo and, then, a
letter by Majorana, where he wrote that he had abandoned his initial suicidal intentions
and decided to return to Naples. However, the following Monday no news from Majorana
reached Carrelli who, worried by these circumstances, called his friend Enrico Fermi in
Rome, who immediately realized the seriousness of the situation: Majorana had
disappeared. At the time, Fermi was working in his laboratory with the young physicist
Giuseppe Cocconi. In order to give him an idea of the serious loss for the community of
physicists caused by Majorana's disappearance, Fermi told Cocconi: ``You see, in the
world there are various categories of scientists: there are people of a secondary or
third level standing, who do their best but do not go very far. There are also those of
high standing, who come to discoveries of great importance, fundamental for the
development of science. Then, there are geniuses like Galileo and Newton. Well, Ettore
was one of them. Majorana had what no-one else in the world had''.

\section{The family background and the first meeting with Fermi}

Ettore Majorana was born on August 5, 1906 in Catania, Sicily (Italy), to Fabio Majorana
and Dorina Corso. The fourth of five sons, he had a rich scientific, technological and
political heritage: three of his uncles were chancellors of the University of Catania and
members of the Italian parliament, while another, Quirino Majorana, was a renowned
experimental physicist and once president of the Italian Physical Society. Ettore's
father himself was an engineer founding the first telephone company in Sicily and later
on a chief inspector of the Ministry of Communications.

Fabio Majorana took care of the education of his son in the first years of his life,
until the family moved to Rome in 1921. Ettore left school in 1923 at the age of 17 and
joined the Faculty of Engineering at the University of Rome, where he became good friend
of Giovanni Gentile Jr and future Nobel laureate Emilio Segr\`e.

After the advent of Fermi to the chair of Theoretical Physics at the University of Rome
in 1926, a group of young people started to form under his guide, supported by Orso Mario
Corbino, the director of the Institute of Physics in Rome and an influential politician,
whose plan of supporting a rapid development of physics in Italy\cite{Segre} led him to
hire Franco Rasetti as his assistant and later on, in spring of 1927, to entice the most
brilliant students of the Faculty of Engineering to move into physics studies. Segr\`e
and his friend Edoardo Amaldi rose to the challenge, joining Fermi and Rasetti's group
and telling them of Ettore's exceptional gifts. After some encouragement from Segr\`e and
Amaldi, Majorana eventually decided to meet Fermi in the autumn of the same year. The two
beautiful minds immediately started talking about the statistical model of atoms that
Fermi was working on, later to be known as the Thomas-Fermi model, which describes the
energy of an atom in terms of the density of its surrounding electrons. The model
involves a complicated non-linear differential equation. The analytical solution of the
equation was unknown at that epoch, but Fermi had managed to obtain a numerical table of
approximate values for it. Majorana carefully followed Fermi reasoning, asked few
questions and left the Institute. The day after he returned to Fermi's office and asked
for a closer look at the numerical table, so that he could compare it with a similar
table he worked up the night before. Once he checked the agreement between the two
results, Majorana said that Fermi's table was correct and left with no further comment.

This anecdote recalled by Rasetti, Segr\`e and Amaldi \cite{Segre,Amaldi} shows that
Majorana arrived at a series solution of the Thomas-Fermi equation using a very peculiar
method. He first transformed the Thomas-Fermi equation into an Abel equation, with a very
original method that can be used for a large class of differential equations; this was
probably done in the sake of applying known theorems on the existence and uniqueness of
solution of the Abel equation. However, he then transformed again the Thomas-Fermi
equation into another first-order differential equation, whose series solution was
explicitly given in terms of only one quadrature, and from this method Majorana obtained
a table of numerical values as accurate as (at least) that of Fermi.

It is a remarkable achievement of Majorana's genius the fact that he obtained many
results well before several renowned mathematicians and physicists or, as for his
solution of the Thomas-Fermi equation (see the historical and scientific assessment
presented in Ref.s \cite{DiGrezia} and \cite{EspositoTF}), which have not been
independently found by anyone else.

\section{Researches in physics}

Majorana gave fundamental contributions in several different areas of theoretical
physics, in part as collaborator of the Fermi group in Rome, but he published only few
scientific articles, so that his brilliant activity was not immediately recognized
outside the Fermi group. The complete list of the published articles is the
following:\footnote{In this list we do not include the short communication presented by
the young student Majorana at the end of 1928, at Fermi's request, during the 1928 annual
meeting of the Italian Physical Society. The published text of that communication
\cite{Nota} was not written by himself; the original work can be found in Ref.
\cite{Volumetti}.}

\begin{enumerate}
\item G. Gentile and E. Majorana, Sullo sdoppiamento dei termini
Roentgen ottici a causa dell'elet\-tro\-ne rotante e sulla
intensit\`a delle righe del Cesio, Rend. Acc. Lincei {\bf 8}, 229
(1928);

\item E. Majorana, Sulla formazione dello ione molecolare di He,
Nuovo Cim. {\bf 8}, 22 (1931);

\item E. Majorana, I presunti termini anomali dell'Elio, Nuovo
Cim. {\bf 8}, 78 (1931);

\item E. Majorana, Reazione pseudopolare fra atomi di Idrogeno,
Rendi. Acc. Lincei {\bf 13}, 58 (1931);

\item E. Majorana, Teoria dei tripletti {\em P'} incompleti, Nuovo
Cim. {\bf 8}, 107 (1931);

\item E. Majorana, Atomi orientati in campo magnetico variabile,
Nuovo Cim. {\bf 9}, 43 (1932);

\item E. Majorana, Teoria relativistica di particelle con momento
intrinseco arbitrario, Nuovo Cim. {\bf 9}, 335 (1932);

\item E. Majorana, \"Uber die Kerntheorie, Z. Phys. {\bf 82}, 137
(1933); Sulla teoria dei nuclei, Ric. Scientifica {\bf 4}(1), 559
(1933);

\item E. Majorana, Teoria simmetrica dell'elettrone e del
positrone, Nuovo Cim. {\bf 14}, 171 (1937);

\item E. Majorana, Il valore delle leggi statistiche nella fisica
e nelle scienze sociali, Scientia {\bf 36}, 55 (1942), edited by
G. Gentile jr. (posthumous).
\end{enumerate}

The largest part of Majorana's work was left unpublished. We are now left with his master
thesis on ``the quantum theory of radioactive nuclei'', 5 notebooks ({\em
``Volumetti''}), 18 booklets ({\em ``Quaderni''}), 12 folders with spare papers, and the
set of the lecture notes on theoretical physics prepared for a class at the University of
Naples. Almost all this material is presently at the Domus Galilaeana in Pisa, Italy; the
complete set of  {\em Volumetti} was translated in English and published in Ref.
\cite{Volumetti}, that of the Naples' lecture notes is (in Italian) in Ref.
\cite{Lezioni}. A forthcoming publication will appear soon (in English) with a reasoned
selection of the most interesting material present in the {\em Quaderni} \cite{Quaderni}.

In the following we give an account of key studies performed by Majorana during his short
scientific life, as emerging from recent historical and scientific researches to date.

\subsection{Atomic and molecular physics}

In 1928, when Majorana started his collaboration (still as a University student in
physics) with the Fermi group in Rome, he already shown an outstanding capacity of
solving very involved mathematical problems in a very interesting and clear way such as,
for example, by obtaining the semi-analytical series solution of the Thomas-Fermi
equation, as mentioned above. The whole work on this topic is contained in some spare
sheets, and diligently reported by the author himself in his {\em Volumetti}
\cite{Volumetti}. From these it is evident the considerable contribution given by
Majorana even in the achievement of the statistical model, anticipating, in many
respects, some results reached later by leading specialists. The major finding by
Majorana was his solution (or, rather, methods of solutions) of the Thomas-Fermi
equation, which remained completely unknown, until recent times \cite{EspositoTF}, to the
physicists community, which failed to realize that the non-linear differential equation
relevant for atoms and other systems could be solved semi-analytically \cite{S7VII}. An
intriguing property in the Majorana derivation of the solution of the Thomas-Fermi
equation is that his method can be easily generalized and applied to a large class of
particular differential equations. Several generalizations of the Thomas-Fermi method for
atoms were proposed as well by Majorana, but they were considered by the physicists
community, unaware of Majorana's unpublished works, only many years later
\cite{DiGrezia}. Indeed, Majorana studied \cite{S16VII} the problem of an atom in a weak
external electric field, i.e., atomic polarizability, and obtained an expression for the
electric dipole moment for a (neutral or arbitrarily ionized) atom. Furthermore, he also
started to consider the application of the statistical method to molecules, rather than
single atoms, studying the case of a diatomic molecule with identical nuclei
\cite{S12VII}. Finally, Majorana also considered the second approximation for the
potential inside the atom, beyond the Thomas-Fermi approximation, with a generalization
of the statistical model of neutral atoms to those ionized $n$ times, including the case
$n = 0$ \cite{S15VII}. This was achieved by focusing the attention on the effective
potential acting on the given electron inside an atom, assumed to be generated by a
rescaled charge density which takes into account the finite charge of the given electron
on which the potential acts. As discussed in Ref. \cite{again}, such an approach is
basically the same which is currently adopted in the renormalization of physical
quantities in modern gauge theories. The specific example considered by Majorana may be
regarded as the first application of this modern viewpoint to an atomic problem.

Some of these works is also recognizable in the underlying framework of the first
published paper 1 made in collaboration with G. Gentile, where they derived the
ionization energy of an electron in the $3d$ orbit of gadolinium and uranium, in good
agreement with the experimental values. In addition, by applying first-order perturbation
theory to the Dirac equation, they also calculated the fine structure splitting of
different (X-ray transitions) spectroscopic terms in gadolinium, uranium and caesium.

After that, by the end of 1931 the 25-year-old physicist had
published two articles 2,4 on the chemical bonding of molecules
and two more papers 3,5 on spectroscopy.

In paper 2, by starting to discuss the still unclear experimental result on the band
structure in helium emission spectrum, Majorana approached the problem of formation of
the molecular ion He$_2^+$ in a way similar to the one proposed earlier by Heitler and
London \cite{HeitlerLondon} for the H$_2$ molecule, where the idea of quantum exchange in
the explanation of the chemical bond was introduced. This same concept based on the
resonance force was also used in paper 4, where he faced the unexplained anomalous $X$
term observed in the spectrum of the H$_2$ molecule, by assuming a pseudopolar binding
between two ions in the hydrogen molecule.

The far-reaching interest in the helium spectrum involved also Majorana who, in paper 3,
performed some calculations on certain doubly excited levels of helium, by taking into
account all helium levels generated by combining two hydrogenoid orbits with principal
quantum number $n=2$ and including the mutual electron repulsion as a perturbation. This
paper anticipated results later obtained (1934-1935) by a collaborator of S. Goudsmith on
the Auger effect in helium and by some other people \cite{Wu}. However, the most
important article published in 1931, dealing with atomic spectroscopy problems, was paper
5, where the characterization of spectra of different atoms with two electrons in the
outer shell was given. A theoretical explanation of the missing lines in the predicted
$P$ triplet levels in the absorption spectra of Hg, Cd and Zn was indeed presented by
introducing a novel process now known as autoionization and equivalent to the Auger
process already known in X-ray emissions. Such a process was independently introduced in
the same year by A.G. Shenstone \cite{Shenstone}, and its important role was later
largely recognized in a great variety of atomic and molecular spectra. The puzzle of the
missing lines was eventually clarified on the experimental side in 1955 with the
observations of W.R.S. Garton and A. Rajaratnam \cite{Garton}, while the correctness of
the Majorana's spectroscopic assignments was proved only in 1970 by W.C. Martin and V.
Kaufman \cite{Martin}.

As Edoardo Amaldi has written \cite{Amaldi}, an in-depth examination of these works
leaves one struck by their quality. They reveal both a deep knowledge of the experimental
data, even in the most minute detail, and an uncommon (and without equal at that time)
ability in using the symmetry properties of the quantum states, resulting in a remarkable
simplification of the problems and a brilliant choice of the most suitable method for
their quantitative resolution.

These published papers, however, do not exhaust the entire work performed by Majorana on
atomic physics, which was the main research topic investigated by the Fermi group in Rome
in the years 1928-1933. On one hand, some echo of the work reported in those papers is
present in the {\em Quaderni} \cite{Quaderni}, where it is pointed out that, especially
in the case of paper 5 on the incomplete $P^\prime$ triplets, some material was not
included by the author in the published work. On the other hand, several other problems
were considered and solved by Majorana, without reporting the results obtained in
published articles, thus being practically unknown to physicists.

Several expressions for the wavefunctions and the different energy levels of two-electron
atoms (and, in particular, of helium) were, for example, considered by Majorana, mainly
in the framework of a variational method aimed at solving the corresponding Schr\"odinger
equation. Numerical values for the energy terms were reported in large summary tables,
and some approximate expressions were also obtained for three-electron atoms (in
particular, for lithium) and for alkali, including the effect of polarization forces in
hydrogen-like atoms \cite{Quaderni}.

The problem of the hyperfine structure of the energy spectra of complex atoms was
considered in some detail as well, revealing the careful attention of Majorana to the
existing literature. A generalization of the Land\`e formula for the hyperfine splitting
to {\em non-Coulomb} atomic field was given, along with a {\em relativistic} formula for
the Rydberg corrections of the hyperfine structures \cite{Quaderni}. Such a detailed
study by Majorana formed the basis of what discussed by Fermi and Segr\`e in a well-known
paper of 1933 on this issue \cite{FermiSegre}, as acknowledged by those authors
themselves.

Some substantial help on difficult theoretical calculations given by Majorana to several
people working in Rome, led in some cases to notable results. For example, in 1932,
stimulated by Segr\`e \cite{Amaldi}, Majorana published his paper 6 on the non-adiabatic
spin-flip of atoms in a magnetic field, which was later extended in 1937 by Nobel
laureate I.I. Rabi and, more in general, in a celebrated work \cite{BlochRabi} of 1945 by
F. Bloch and Rabi (who explicitly recognized the work by Majorana as seminal for the
solution of the problem). It established the theoretical basis for the experimental
method used to reverse the spin of neutrons by a radio-frequency field, a method that is
still employed today in all polarized-neutron spectrometers. That paper contained also an
independent derivation of the well-known Landau-Zener formula (published later in the
same year) \cite{LandauZener} for non-adiabatic transition probability. It also
introduced a novel mathematical tool for representing spherical functions (Majorana
sphere), rediscovered only in recent times.

Several problems of molecular physics were also faced by Majorana. He studied in some
detail, for example, the helium molecule and considered the general theory of the
vibration modes in molecules as well, with particular reference to the C$_2$H$_2$
molecule of acetylene (which presents peculiar geometric properties) \cite{Quaderni}.

Finally, other notable results concerned the problem of ferromagnetism in the framework
of the Heisenberg model with the exchange interaction. The approach used by Majorana in
this study, however, is {\it original} \cite{Quaderni}, since it does not follow neither
the Heisenberg formulation \cite{HeisenbergFerro} nor the subsequent van Vleck
formulation \cite{Vleck} in terms of spin Hamiltonian. By using statistical arguments, he
calculated the magnetization (with respect to the saturation value) of the ferromagnetic
system when an external magnetic field is applied, and the spontaneous magnetization.
Several examples of ferromagnetic materials, with different geometries, were also
reported in his own notebooks.

\subsection{Nuclear physics}

Majorana was among the firsts who studied nuclear physics in Rome, at least since 1929
when he defended his master thesis on ``the quantum theory of radioactive nuclei''.
However quite unexpectedly, since the Fermi group was not effectively involved in nuclear
physics until 1933, he continued to study such topics for several years, independently of
the main research topics of the Fermi group, till his famous theory of nuclear exchange
forces published in paper 8 of 1933. The antecedents that led to such a discovery are
quite intriguing and we dwell here a bit on these.

In March 1932 James Chadwick announced the discovery of the neutron, after which Majorana
revealed to his friends and colleagues in Rome \cite{Amaldi} that he had built a theory
of light nuclei based on the quantum concept of exchange forces. Although encouraged by
Fermi to go public with his results, Majorana's hypercritical judgement prevented him
from doing so. True to style, his work was not recognized until a few months later when
it was independently elaborated by Werner Heisenberg. This fact caused a sensation in the
Rome group, and Fermi urged Majorana, successfully, to visit Heisenberg in Leipzig, for a
six-month period in 1933, where Majorana's capacities and results largely impressed
Heisenberg himself.

In the Heisenberg model, atomic nuclei were supposed to be composed of protons and
neutrons only, without any need of electrons as was previously largely accepted (before
the discovery of the neutron, protons and electrons were the only known ``elementary''
particles). It was also assumed that the leading forces responsible for nuclear stability
were those between neutrons and protons (neglecting neutron-neutron and proton-proton
forces), which were deduced to be exchange forces in analogy to what had been proven by
Heitler and London for the H$_2^+$ molecular ion (in this case, an H atom and an H$^+$
ion are held together by the exchange of an electron). According to Heisenberg, the
underlying nuclear forces should be interpreted in terms of nucleons exchanging spinless
electrons, implicitly assuming that the neutron was practically formed by a proton and an
electron. Majorana immediately realized this defect of the Heisenberg's theory. Indeed,
in Majorana's view, the neutron was pictured as a ``neutral proton" \cite{Amaldi}, as
effectively is, and in his model the forces between neutrons and protons were explained
in terms of the exchange interaction arising from the quantum effect coming from
interchanging the space coordinates of identical nucleons, rather than from interchanging
their electric charge as in the Heisenberg model. The direct consequences of the Majorana
version of the nuclear model were that the exchange forces considered are independent of
nuclear density of the total mass and prevent the collapse of the nucleus even without a
repulsive force at short distance. The observed particular stability of the
$\alpha$-particle (and not the unobserved one of deuterium nucleus, as predicted by the
Heisenberg model) was thus explained by a ``saturation phenomenon more or less analogous
to valence saturation''. It was especially this saturation of nuclear forces for the
$\alpha$-particle that quickly led to the recognition by Heisenberg himself and others of
the Majorana model as the most appropriate one. On several occasions, in fact, while
discussing the ``Heisenberg-Majorana'' exchange forces, Heisenberg mentioned his own
contribution only marginally, rather emphasizing that of Majorana \cite{DeGregorio}.

This model was certainly the most renowned contribution by Majorana to the contemporary
physicists community; however, he did not limit himself to study this particular nuclear
physics topic. In the research notes of the {\em Quaderni} \cite{Quaderni}, several pages
were devoted to study possible forms of the nucleon potential inside a given nucleus,
describing the interaction between neutrons and protons. Although generic nuclei were
often considered in the discussion, some particular care was given by Majorana to light
nuclei (deuteron, $\alpha$-particles, etc.). In a sense, the researches performed by
Majorana on this subject are, at the same time, preliminary studies and generalizations
of what had been published by himself in his well-known paper 8, revealing a very rich
and peculiar way of reasoning. Note also that, probably before his studies leading to
paper 8, a relativistically invariant field theory for nuclei composed of scalar
particles was also elaborated by Majorana \cite{PauliWeisskopf}, which described the
transitions between different nuclei.

In 1930 Majorana elaborated a dynamical theory of $(\alpha,p)$ reactions on light nuclei
\cite{S28VIV}, whose experimental results were interpreted by Chadwick and G. Gamov
\cite{ChadGamov}, describing the energy states in terms of the superposition of a
continuous spectrum and a discrete level. He provided a complete theory for the
artificial disintegration of nuclei by means of $\alpha$-particles (with and without
$\alpha$ absorption) approaching the problem by considering the simplest case with an
unstable state of the system formed by a nucleus plus an $\alpha$-particle, which
spontaneously decays with the emission of an $\alpha$-particle or a proton. In particular
Majorana obtained the explicit expression for the integrated cross section of the nuclear
process. The peculiar aspect of Majorana theory \cite{quasistat} was the introduction of
quasi-stationary states (the superposition of discrete and continuous states), which were
only later considered by U. Fano \cite{Fano} in 1935 in a completely different context
and then widely applied in the framework of condensed matter physics about 20 years
later.

In addition to this, other topics were also considered by Majorana
\cite{Volumetti,Quaderni}, and we here only mention the study of the problem of the
energy loss of $\beta$-particles passing through a medium, where he deduced the Thomson
formula by using classical arguments. This study could be potentially of some interest,
for correct historical reconstructions, in relation with the famous theory elaborated by
Fermi on the nuclear $\beta$ decay just in 1934.

\subsection{Relativistic fields and group theory}

Among the papers published by Majorana in 1932, the most important one is certainly paper
7 concerning a relativistic wave mechanics of particles with arbitrary spin, which makes
no use of the negative-energy states. Around 1932 it was commonly thought that one could
write relativistic quantum equations only in the case of particles with zero or
half-integer spin. Majorana had quite a different belief (see several works carried out
in the {\em Quaderni} \cite{Quaderni}), and he began constructing suitable
quantum-relativistic equations for higher spin values (one, three-halves, etc.), even
devising a general method for writing the equation for a generic spin-value. Still, he
did not publish anything, until he discovered that one could write a single equation to
cover an infinite series of cases, that is, a whole, infinite family of particles of
arbitrary spin: this was finally reported in paper 7. In order to implement his programme
with these ``infinite components" equations, Majorana invented a technique for the
representation of a group several years before Eugene Wigner did. Remarkably, Majorana
obtained the simplest infinite-dimensional unitary representations of the Lorentz group
that were re-discovered by Wigner in his 1939 and 1948 works \cite{Wigner}. The entire
theory was re-invented by Soviet mathematicians (in particular Gelfand and collaborators
\cite{Gelfand,GelfandRev}) in a series of articles starting from 1948 and finally applied
by physicists, first to hadronic physics and then to modern string theories, years later.
The importance of Majorana's article was first realized by B.L. van der Waerden
\cite{Waerden} but, unfortunately, it remained unnoticed for more than three decades
until D. Fradkin, informed by Amaldi, reexamined that pioneering work in light of later
developments and clearly explained the relevance of Majorana's approach and results
accomplished many years earlier \cite{Fradkin}.

Some work behind paper 7 can be found as well in the {\em
Quaderni} \cite{Quaderni}. Here, by starting from the usual Dirac
equation for a 4-component spinor, he obtained explicit
expressions for the Dirac matrices in the cases of 6-component and
16-component spinors. Interesting enough, at the end of his
discussion, Majorana also treated the case of spinors with an {\it
odd} number of components, namely a 5-component field.

Majorana's admiration for the work of Hermann Weyl and others on the application of group
theory to quantum mechanics was first recalled by Amaldi \cite{Amaldi} and, apart from
the published paper 7, is well testified by a number of unpublished notes
\cite{Volumetti,Quaderni}. The Weyl approach, indeed, greatly influenced the entire
scientific thought and work of Majorana \cite{DragoEspo}. This is particularly evident in
his notebooks \cite{VIIIVV} where, for example, he gave a detailed analysis of the
relationship between the representations of the Lorentz group and the matrices of the
(special) unitary group in two dimensions. In these notes, a strict connection with the
Dirac equation was always taken into account, and the explicit form of the
transformations of every bilinear in a given spinor field which is relevant in the Dirac
theory of a spinning particle was reported.

The Dirac equation was, indeed, a favorite and long-lasting topic studied by Majorana. In
the {\em Quaderni} \cite{Quaderni}, the relativistic equation describing spin-1/2
particles was usually considered in a Lagrangian framework (in general, the canonical
formalism was adopted), obtained from a least action principle. After an interesting
preliminary study of the problem of the vibrating string, where Majorana obtained a
(classical) Dirac-like equation for a two-component field, he then went on to consider a
semiclassical relativistic theory for the electron, where the Klein-Gordon equation and
the Dirac equation were deduced from a semiclassical Hamilton-Jacobi equation. Later on,
the field equations and their properties (Lorentz invariance, issues related to the
probabilistic interpretation, and so on) were considered in details, and the quantization
of the (free) Dirac field discussed by means of the standard formalism, using
annihilation and creation operator formalism. Finally, the electromagnetic interaction
was introduced in the Dirac equation and the superposition of the Dirac and Maxwell
fields studied in a very peculiar way, obtaining the expression for the quantized
Hamiltonian of the interacting system using a normal mode decomposition.

Real (rather than complex) Dirac fields were considered by Majorana in his paper 9 on a
``symmetrical theory for electrons and positrons'', where he introduced the so-called
(and now well-known to particle physicists) Majorana neutrino hypothesis. This hypothesis
was no doubt revolutionary, because it first put forward the possibility that the
antimatter partner of a given matter particle could be the particle itself. This was in
direct contradiction to what Dirac had successfully assumed in order to solve the problem
of negative-energy states in quantum field theory (i.e., the existence of the positron).
With amazing farsightedness Majorana suggested that the neutrino, which had just been
postulated by Wolfgang Pauli and Fermi to explain puzzling features of radioactive beta
decay, could be such a particle. This would make the neutrino unique among the elementary
particles and, moreover, enable it to have mass. Today many experiments are still devoted
to detect these peculiar properties, which include the phenomenon of neutrino
oscillations: we have not yet succeeded to find a definite answer to Majorana's proposal.

\subsection{Quantum electrodynamics}

A large part of Majorana's interests was devoted to several theoretical problems of the
rising quantum electrodynamics \cite{Volumetti,Quaderni}, especially after he returned
from Leipzig (after 1933). This is also testified by a letter to his uncle Quirino
\cite{Quirino}, who constantly pressed Majorana for theoretical explanations of his own
experiments \cite{Dragoni}. However, also in this case, according to his hypercritical
judgment Majorana decided not to publish any paper on his studies.

Fortunately we know from his notebooks that he generally considered quantum
electrodynamics in a Lagrangian and Hamiltonian framework, with the use of a least action
principle. In one of his studies \cite{Quaderni}, as it is {\it now} customary, the
electromagnetic field was decomposed in plane wave operators, and its properties studied
in a {full Lorentz-invariant formalism} by employing group-theoretic arguments. Explicit
expressions for the quantized Hamiltonian, creation and annihilation operators for the
photons, as well as angular momentum operator, were deduced in several different bases,
along with the appropriate commutation relations. Apart from the relevance of the
scientific results contained in such writings, and achieved by Majorana well in advance
with respect to other scholars, they also testify on the approach followed by Majorana on
dealing with these topics. In fact, once more being ahead of his time, he adopted the
more mathematical and more powerful methods introduced by Heisenberg, Born, Jordan and
Klein, instead of following the plain (and quite famous) approach by Fermi \cite{Miller}.

This is also the case for a recently retrieved text written by
Majorana in French \cite{Francese}, where Majorana dealt with
quite a peculiar topic in quantum electrodynamics. It is
instructive, for this topic, just to quote directly from the
Majorana's paper:
\begin{quote}
Let us consider a system of $p$ electrons and put the following
assumptions: 1) the interaction between the particles is
sufficiently small allowing to speak about individual quantum
states, so that we may consider that the quantum numbers defining
the configuration of the system are good quantum numbers; 2) any
electron has a number $n>p$ of inner energetic levels, while any
other level has a much greater energy. We deduce that the states
of the system as a whole may be divided into two classes. The
first one is composed of those configurations for which all the
electrons belong to one of the inner states. Instead the second
one is formed by those configurations in which at least one
electron belongs to a higher level not included in the n levels
already mentioned. We will also assume that it is possible, with a
sufficiently degree of approximation, to neglect the interaction
between the states of the two classes. In other words we will
neglect the matrix elements of the energy corresponding to the
coupling of different classes, so that we may consider the motion
of the p particles in the n inner states, as if only these states
exist. Then, our aim is to translate this problem into that of the
motion of $n-p$ particles in the same states, such new particles
representing the holes, according to the Pauli principle.
\end{quote}
Majorana, thus, by following a track left by Heisenberg, applied the formalism of field
quantization to Dirac's hole theory, obtaining the general expression for the QED
Hamiltonian in terms of anticommuting holes quantities. We also point out the peculiar
justification for using anticommutators for fermionic variables given by Majorana; this,
indeed, ``cannot be justified on general grounds, but only by the particular form of the
Hamiltonian. In fact, we may verify that the equations of motion are satisfied to the
best by these last exchange relations rather than by the Heisenberg ones''.

In the {\em Quaderni} we also find several studies, following an
original idea by Oppenheimer \cite{Oppenheimer}, aimed at
exploring the possibility to describe the electromagnetic field by
means of the Dirac formalism, as already pointed out in the
literature. Some emphasis was given to consider the properties of
the electromagnetic field as described by a real wavefunction for
the photon, a study which was well beyond the contemporary works
of other authors \cite{Real}.

Another unknown contribution by Majorana, which has been brought to light only recently
\cite{PauliWeisskopf} is nothing but the theory of scalar quantum electrodynamics, as
elaborated several years later (1934) by Pauli and V.F. Weisskopf \cite{PW}. The results
obtained by Majorana were well summarized by Gian Carlo Wick \cite{Wick}, who was aware
of them directly from Majorana (although Wick never saw the written calculations,
reported in \cite{Quaderni}):
\begin{quote}
[...] I was admitted as a bystander to an international meeting on
nuclear physics, which was organized by the Royal Academy in Rome
in the fall of 1931. [...]

I was asked by Heitler to act as a sort of interpreter between him and Majorana. He spoke
hardly any Italian, and Majorana's German was a bit weak. So during lunches, Heitler
expressed a curiosity in what Majorana was doing; Fermi must have told him how bright he
was. So Majorana began telling, in that detached and somewhat ironical tone, which was
typical of him, especially when discussing his own work, that he was developing a
relativistic theory for charged particles. It was not true, he said, that the
Schr\"odinger equation for a relativistic particle had to have the form indicated by
Dirac. It was clear by now, that in a relativistic theory one had to start from a field
theory and for this purpose the Klein-Gordon wave-equation was just as legitimate as
Dirac's. If one starts with the first one, and quantizes it, one get a theory with
consequences quite similar to Dirac positron theory, with positive and negative charges,
the possibility of pair creation etc. You can see what I am driving at: Majorana had the
Pauli-Weisskopf scheme all worked out already at that time. Please do not think I am
disputing the merit of these authors. Majorana never published this work, he did not seem
to take it very seriously, and I don't think Pauli and Weisskopf ever even heard of it.
Heitler probably forgot all about it, and so did I, until I saw the paper by Pauli and
Weisskopf.
\end{quote}
As well known, the relevance of the Pauli-Weisskopf theory for a scalar field is
certainly related to its possible direct applications, i.e. meson theory, but, in
addition, it can be considered a milestone in the development of quantum electrodynamics.
In fact, it showed unambiguously that the marriage between quantum mechanics and special
relativity did not necessarily require a spin 1/2 for the correct interpretation of the
formalism, as erroneously believed. In this respect, it is interesting that the
formulation of such a theory by Majorana can be dated as early as 1929-1930. The Majorana
theory, however, presents also several different theoretical aspects with respect to the
Pauli-Weisskopf theory: this is the case, for example, for the use of {\it general} sets
of plane waves in the expansion of the field variables, or the adoption of four (for
matter particles) plus four (for photons) instead of four plus two operators describing
the quanta of the appropriate fields.

Other peculiar investigations \cite{Quaderni} concerned the possibility to introduce an
{intrinsic} time delay (as a universal constant) in the expressions for the
electromagnetic retarded fields, and the study on the modification of the Maxwell
equations in presence of magnetic monopoles. Besides the fully theoretical work about
quantum electrodynamics, some applications were also considered by Majorana. This is the
case of the free electron scattering, where he gave an {explicit} expression for the
transition probability, and the coherent scattering of bound electrons. Several other
scattering processes were considered as well \cite{Quaderni} in the framework of
perturbation theory, by means of the Dirac or the Born methods.

Finally, we also mention an intriguing attempt, made by Majorana as early as in 1928, to
find a relation between the fundamental constants $e, h, c$ \cite{S30VII}. The interest
for this work is not related to the particular mechanical model used by him (that,
indeed, led to a result $e^2 \simeq h c$ far from the truth, as noted by Majorana
himself), but rather to the interpretation of the electromagnetic interaction in terms of
exchanged particles. According to Majorana, indeed, the electromagnetic field generated
by charged particles is, in some sense, quantized, and two electrons interact by
exchanging particles each another.
\begin{quote}
As a first approximation we can describe the situation in terms of
a pointlike mass moving with a group velocity equal to the light
speed $c$. Let us also make the arbitrary assumption that such
point-particle moves periodically between $A$ and $B$ and back.
Let us further suppose that it is free of interactions while
travelling between $A$ and $B$, whereas in $A$ and in $B$ it
inverts its velocity due to the collision with the electrons
setting there.
\end{quote}
As it is evident, such an interpretation of the quantized
electromagnetic field substantially coincides with that introduced
more than a decade later by Feynman in quantum electrodynamics,
the point-particles exchanged by electrons being assumed to be the
photons.

\subsection{Other studies}

In some notes, probably prepared for a seminar at the University
of Naples in 1938 \cite{path}, Majorana gave a physical
interpretation of quantum mechanics which anticipated of several
years the Feynman approach in terms of path integral,
independently of the underlying mathematical formulation.

The starting point in Majorana's paper was to search for a meaningful and clear
formulation of the concept of quantum state. This was achieved by considering some sets
of `solutions that differ for the initial conditions' which, in the Feynman language of
1948 \cite{Feynman}, correspond precisely to the different integration paths: the
different initial conditions were always referred to the same initial time, while the
determined quantum state corresponded to a fixed end time. Moreover, the crucial point in
the Feynman formulation of quantum mechanics, namely that of considering not only the
paths corresponding to classical trajectories, but all possible paths joining the initial
and final points, was introduced by Majorana after an interesting discussion on the
harmonic oscillator, where its quantum energy levels were interpreted in terms of
classical oscillations: ``we can say that the ground state with energy $E_0 = h \nu /2$
corresponds roughly to all classical oscillations with energy between 0 and $h \nu$, the
first excited state with energy $E_1 = 3 h \nu /2$ corresponds to the classical solutions
with energy between $h \nu$ and $2 h \nu$, and so on.'' Here we stress also the key role
played by the symmetry properties of the physical system in Majorana analysis, a feature
which is, again, quite common in all his papers. Furthermore, although no trace can be
found of the formalism underlying the Feynman path-integral approach to quantum mechanics
in the Majorana manuscript, nevertheless it is very intriguing that the main physical
issue, the novel way of interpreting the theory of quanta, were realized well in advance
by Majorana. This is particularly impressive if we take into account that, in the known
historical path, the interpretation of the formalism has only followed the mathematical
development of the formalism itself.

Finally, we also mention paper 10, published posthumous by Majorana's friend Gentile and
dealing with a completely different subject: ``the value of statistical laws in physics
and social sciences''. As suggested by Gentile, this was probably intended for a
sociology journal, and was likely written during a period ranging from 1933 to 1937, when
Majorana's studies extended to cover economics, politics, philosophical problems, etc.
This paper presented the point of view of a physicist about the value of statistical
laws, in physics and social sciences, to people of different disciplines such as
sociology and economics. Of some interest was the observation that statistical laws are
investigation tools to be used in economic and social modelling, and have the same
epistemological status of irreducible probabilistic laws as quantum mechanics. Then, even
in this case, the approach by Majorana was well ahead with respect to that of the vast
majority of his contemporaries working in this field.

\section{Professor of theoretical physics}

As we have seen, Majorana contributed significantly to many theoretical researches which
were considered as the frontier topics in the 1930s. However, his own peculiar
contribution ranged also on the basics and applications of quantum mechanics, as
Majorana's lectures on theoretical physics clearly demonstrate.

As realized only recently, Majorana revealed a genuine interest in advanced physics
teaching starting from 1933, soon after he obtained at the end of 1932 the professorship
degree of ``libero docente'' (analogous to the German privatdozent). In view of this
position, he proposed some academic courses at the University of Rome, as testified by
the programs of three courses he would have given between 1933 and 1937 \cite{Teaching}.
Although Majorana never gave lectures, probably due to the lacking of students, they are
particularly interesting and informative due to a very careful choice of the topics he
intended to treat in his courses.

The first course (academic year 1933-34) was that of Mathematical Methods of Quantum
Mechanics, whose syllabus contained the following topics:
\begin{quote}
1) Unitary geometry. Linear transformations. Hermitian operators.
Unitary transformations. Eigenvalues and eigenvectors. 2) Phase
space and the quantum of action. Modifications to classical
kinematics. General framework of quantum mechanics. 3)
Hamiltonians which are invariant under a transformation group.
Transformations as complex quantities. Non compatible systems.
Representations of finite or continuous groups. 4) General
elements on abstract groups. Representation theorems. The group of
spatial rotations. Symmetric groups of permutations and other
finite groups. 5) Properties of the systems endowed with spherical
symmetry. Orbital and intrinsic momenta. Theory of the rigid
rotator. 6) Systems with identical particles. Fermi and
Bose-Einstein statistics. Symmetries of the eigenfunctions in the
center-of-mass frames. 7) The Lorentz group and the spinor
calculus. Applications to the relativistic theory of the
elementary particles.
\end{quote}
The second course (academic year 1935-36) was instead on Mathematical Methods of Atomic
Physics covering the following topics:
\begin{quote}
Matrix calculus. Phase space and the correspondence principle.
Minimal statistical sets or elementary cells. Elements of the
quantum dynamics. Statistical theories. General definition of
symmetry problems. Representations of groups. Complex atomic
spectra. Kinematics of the rigid body. Diatomic and polyatomic
molecules. Relativistic theory of the electron and the foundations
of electrodynamics. Hyperfine structures and alternating bands.
Elements of nuclear physics.
\end{quote}
Finally, the third course (academic year 1936-37) was planned to discuss Quantum
Electrodynamics:
\begin{quote}
Relativistic theory of the electron. Quantization procedures. Field quantities defined by
commutability and anticommutability laws. Their kinematical equivalence with sets with an
undetermined number of objects obeying the Bose-Einstein or Fermi statistics. Dynamical
equivalence. Quantization of the Maxwell-Dirac equations. Study of the relativistic
invariance. The positive electron and the symmetry of charges. Several applications of
the theory. Radiation and scattering processes. Creation and annihilation of opposite
charges. Collisions of fast electrons.
\end{quote}

Majorana effectively lectured on theoretical physics only in 1938 when, as recalled
above, he obtained a position as a full professor at University of Naples. He started on
January 13 till March 24, one day before his disappearance, thus about two months only,
but his activity was intense, and his interest for teaching extremely high. For the
benefit of his students, he prepared careful notes for his lectures \cite{Lezioni}. A
recent analysis has shown that Majorana's 1938 course was very innovative for that time
\cite{Teaching,DragoEspo}, and this has been confirmed by the retrieval (on September
2004) of a faithful transcription of the whole set of Majorana's lecture notes (the
so-called ``Moreno lecture notes'') comprising 6 lectures not included in the original
collection \cite{Moreno}.

The first part of his course on theoretical physics dealt with the phenomenology of the
atomic physics and its interpretation in the framework of the old quantum theory of
Bohr-Sommerfeld. This part was quite analogous to the course given by Fermi in Rome
(1927-28) attended by the student Majorana.

The second part started with the classical radiation theory, reporting explicit solutions
of the Maxwell equations, scattering of the solar light and some other applications. It
then discussed the theory of relativity: after the presentation of the corresponding
phenomenology, a complete discussion of the mathematical formalism required by the theory
was given, ending with some applications as the relativistic dynamics of the electron.
Finally, a discussion of several effects for the interpretation of quantum mechanics,
such as the photoelectric effect, the Thomson scattering, the Compton effects and the
Franck-Hertz experiment, were addressed.

The last part of the course, more mathematically oriented, dealt with the
Schr\"o\-din\-ger and the Heisenberg formulation of quantum mechanics. This part did not
follow the Fermi approach, but rather referred to previous personal studies by Majorana
\cite{Volumetti}, also following the original Weyl's book \cite{Weyl} on group theory and
quantum mechanics.

Majorana's course(s) thus consisted of a fruitful mixture of an original approach - very
similar to that of today's courses on quantum mechanics - and of some consolidated lines
of development, which were the clear legacy of the lines Fermi had adopted in his
courses. Both the lecture notes for the Naples' course and the programs of the three
courses that Majorana submitted in Rome between 1933 and 1936, reveals his forefront
approach and his search for new routes in dealing with quantum mechanics.

\section{Conclusion}

Majorana's course on theoretical physics was suddenly and
unexpectedly interrupted by his mysterious disappearance three
months after his appointment at Naples.

On Friday March 25, 1938 Majorana went to the Institute of Physics and handed over the
lecture notes and some other papers to one of his students. After that, he returned to
his hotel and, after writing farewell letters to his family and to the director of the
Institute of Physics, Carrelli, apparently embarked on a ship to Palermo. He reached his
destination the following morning, where he lodged for a short time in the Grand Hotel
Sole. It was there that he wrote a telegram and a letter to Carrelli pointing out a
change of mind about his decisions. On Saturday evening Majorana embarked on a ship from
Palermo to Naples. From here onwards, no other reliable information about him are
available.

There have been several conjectures about the fate of Majorana, including suicide, a
retreat in a monastery and a flight to a foreign country \cite{Recami}. Understanding the
root of such a dramatic decision is perhaps impossible and it could be triggered by
personal and familiar reasons, such as Majorana's peculiar relationship with his
extremely possessive mother (especially after the death of his father in 1934), or more
elaborated reasons reported in nice literary tales. However, quoting Majorana himself on
his approach to physics: ``We cannot give to such hypothesis greater likelihood than to
some other theoretical presumptions without a too much subjective appraisal''.

Attention should not, however, be shifted far from the outstanding
work performed by the Italian scientist, only briefly outlined in
this review. We prefer to end with Fermi's own words, written not
long after Majorana's disappearance \cite{Duce}:
\begin{quote}
Able at the same time to develop audacious hypothesis and
criticize acutely his work and that of others; very skilled
calculating man, a deep-routed mathematician that never loses the
very essence of the physical problem behind the veil of numbers
and algorithms, Ettore Majorana has at the highest level that rare
collection of abilities which form the theoretical physicist of
very first-rank. Indeed, in the few years during which his
activity has been carried out, until now, he has been able to
outclass the attention of scholars from all over the world, who
recognized, in his works, the stamp of one of the greatest mind of
our times and the promise of further conquests.
\end{quote}


\subsection*{Acknowledgments}
The interest of the present author in the work by Ettore Majorana was stimulated, many
years ago, by Erasmo Recami who, since then, always encouraged him in further studies.
His help and support is here gratefully acknowledged. The author is also indebted to
Gianpiero Mangano for a very careful reading of the manuscript.

\newpage
\appendix

\section{Chronology of the life and work of Ettore Majorana}


\noindent {\bf 1906}
\\ 
\noindent {\it August 5} \\
Ettore Majorana was born in Catania to Fabio Massimo (1875-1934)
and Salvatrice (Dorina) Corso (1876-1965), fourth of five sons
(Rosina, Salvatore, Luciano, Ettore and Maria).
\\ ${}$

\noindent {\bf 1921}
\\ 
\noindent The Majorana family moved from Sicily to Rome.
\\ ${}$

\noindent {\bf 1923}
\\ 
\noindent {\it July} \\
He completed secondary school (Liceo-Ginnasio Statale ``Torquato Tasso'' in Rome) at the
age of 17.
\\ 
{\it November 3} \\
He joined the Faculty of Engineering at the University of Rome.
\\ ${}$

\noindent {\bf 1927}
\\ 
\noindent He started to write the ``Volumetti'', personal notebooks where he wrote down
his own studies and/or researches (the date reported in the first of the five notebooks
is March 8, 1927).
\\ 
{\it Summer} \\
E. Amaldi and E. Segr\`e, students in Engineering, decided to join the Fermi group (the
formal passage was registered on November 1927 for Segr\`e and on February 1928 for
Amaldi), on advice of the Director of the Institute of Physics in Rome, O.M. Corbino.
\\ 
{\it Autumn} \\
Segr\`e encouraged Majorana to meet Fermi; after this meeting, he moved from Engineering
to Physics studies(the formal passage was registered on November 19, 1928).
\\ ${}$

\noindent {\bf 1928}
\\ 
\noindent In collaboration with G. Gentile jr. he published his
first paper 1, following the spectroscopic researches of the Fermi
group (the paper was presented at the Accademia dei Lincei on July
24).
\\ 
{\it December 29} \\
Still University student, he participated at the XXII General Meeting of the Italian
Physical Society (directed by his uncle Quirino), giving a talk on a {\it Ricerca di
un'espressione generale delle correzioni di Rydberg, valevole per atomi neutri o
ionizzati positivamente} (the summary of his talk appeared in Nuovo Cim. {\bf 6}, XIV
(1929) while the original work is in the Volumetto II.)
\\ ${}$

\noindent {\bf 1929}
\\ 
\noindent {\it July 6} \\
He graduated with a master degree in Physics, with a thesis on {\it La teoria quantistica
dei nuclei radioattivi} ({\it The quantum theory of radioactive nuclei}). It is the first
time that the subject of nuclear physics appears among the activity of the Rome group.
\\ ${}$

\noindent {\bf 1931}
\\ 
\noindent He published two papers (2,4) on the chemical bond of molecules and two further
papers (3,5) on spectroscopic researches.
\\ ${}$

\noindent {\bf 1932}
\\ 
\noindent He published two papers (6,7). In the first one, stimulated by Segr\`e, he
studied for the first time, from a theoretical point of view, the non-adiabatic
spin-flip, whose transition probability was independently obtained by Landau and Zener in
the same year.
In the second paper, again for the first time, he introduced the infinite-dimensional
representation of the Lorentz group, anticipating the results obtained by E. Wigner in
1938 and 1948.
\\ 
{\it March} \\
After the discovery of the neutron by J. Chadwick, Majorana discuss with the members of
Fermi group a theory of light nuclei made of only protons and neutrons (without
electrons). Although encouraged by Fermi and his group, he decided to not publish his
work. A similar result, with some imperfection, is published independently by W.
Heisenberg in the following July.
\\ 
{\it November 12} \\
He got the ``Libera Docenza'' degree in Theoretical Physics at the University of Rome.
\\ ${}$

\noindent {\bf 1933}
\\ 
\noindent {\it January} \\
Upon suggestion of E. Fermi, he obtained a fellowship from the Italian C.N.R. for
visiting the Institute of Theoretical Physics in Leipzig head by Heisenberg.
\\ 
{\it January 19} \\
In the evening he arrived at Leipzig.
\\ 
{\it March 3} \\
Encouraged by Heisenberg, he finally sent, to the German journal Zeitschrift f\"ur
Physik, his paper 8 on nuclear theory, whose main result had been already obtained one
year before and improving the Heisenberg theory on nuclear interactions.
\\ 
{\it March 4} \\
From Leipzig he moved to Copenhagen, where he stayed for about one month at the Institute
of Theoretical Physics directed by N. Bohr.
\\ 
{\it April 15} \\
He left Copenhagen and come back to Rome for the Easter holidays.
\\ 
{\it May} \\
At the University of Rome he presented the syllabus of a course on Mathematical Methods
of Quantum Mechanics, which he never taught.
\\ 
{\it May 5} \\
He returned to Leipzig to continue his visit.
\\ 
{\it May 11} \\
Upon request of the Italian C.N.R., he sent to the official journal of this agency the
Italian version of the paper 8 published in German.
\\ 
{\it July} \\
The Majorana family (the mother of Ettore with his sisters Rosina and Maria and
his brother Salvatore) visited Ettore in Leipzig.
\\ 
{\it August 5} \\
He definitively returned to Rome.
\\ ${}$

\noindent {\bf 1934}
\\ 
\noindent His father died; such a tragic event had a strong influence on his following
life.
\\ ${}$

\noindent {\bf 1935}
\\ 
\noindent {\it April 30} \\
At the University of Rome he presented the syllabus of a course on Mathematical Methods
of Atomic Physics, he never taught.
\\ ${}$

\noindent {\bf 1936}
\\ 
\noindent {\it April 28} \\
At the University of Rome he presented the syllabus and materials for a course on Quantum
Electrodynamics, never taught in the following.
\\ ${}$

\noindent {\bf 1937}
\\ 
\noindent He published his last paper 9, where he presented results obtained some years
before. It contained the fundamental theory on the Majorana neutrino, which twenty years
later has been recognized as a basic result in order to explain the phenomena related to
the problem of the neutrino mass.
\\ 
{\it June} \\
He decided to participate to the national call for a full professorship in Theoretical
Physics at the University of Palermo (obtained at the request of E. Segr\`e). Among the
participants we find G. Gentile jr, L. Pincherle, G. Racah, G. Wataghin, G.C. Wick.
\\ 
{\it October 25} \\
The committee for the competition (headed by Fermi) met for the first time, and
immediately stopped its work in order to send a letter to the Minister of Education with
the request to give a professorship in Theoretical Physics to Majorana ``for high and
well deserved repute'', independently of the usual competition rules.
\\ 
{\it November 2} \\
The Minister of Education accepted the request and appointed Majorana as full professor
at the University of Naples (starting from the following November 16).
\\ ${}$

\noindent {\bf 1938}
\\ 
\noindent {\it January 10 or 11} \\
He arrived in Naples to take up the chair of Theoretical Physics.
\\ 
{\it January 13} \\
At 9.00 he delivered the opening lecture for his course on Theoretical
Physics. His family arrived from Rome for the occasion.
\\ 
{\it January 15} \\
He effectively started his course (which was usually delivered on Tuesday, Thursday and
Saturday of each week).
\\ 
{\it March 12} \\
After his lecture, he visited his family in Rome for the last time.
\\ 
{\it March 24} \\
He gave his last lecture (N.21).
\\ 
{\it March 25} \\
In the morning he went to the Institute of Physics of Naples to give a folder with the
notes of his lectures to a student of him. After a stop at his Hotel, at 17.00 he left
for an unknown destination.

${}$\\

\end{document}